\documentclass[twocolumn,10pt,aps,amsmath,amssymb,nofootinbib,superscriptaddress,showkeys,showpacs]{revtex4-1}
\usepackage{epsfig} 
\usepackage{amsmath} 
\usepackage{graphicx}
\usepackage{color}
\usepackage{float}
\newcommand{\ba}{\begin{array}}
\newcommand{\ea}{\end{array}}
\newcommand{\bea}{\begin{eqnarray}}
\newcommand{\eea}{\end{eqnarray}}

\newcommand{\be}{\begin{equation}}
\newcommand{\ee}{\end{equation}}

\def\bra#1{\left\langle #1\right|}
\def\ket#1{\left| #1\right\rangle}
\begin{document} 

\title{Re-examining $\sin 2 \beta$ and $\Delta m_d$  from evolution of $B^0_d$ mesons with decoherence}

\author{Ashutosh Kumar Alok}
\email{akalok@iitj.ac.in}
\affiliation{Indian Institute of Technology Jodhpur, Jodhpur 342011, India}

\author{Subhashish Banerjee}
\email{subhashish@iitj.ac.in}
\affiliation{Indian Institute of Technology Jodhpur, Jodhpur 342011, India}

\author{S. Uma Sankar}
\email{uma@phy.iitb.ac.in}
\affiliation{Indian  Institute  of  Technology  Bombay,  Mumbai  - 400076, India}

\begin{abstract}
In the time evolution of neutral meson systems, a perfect quantum coherence is usually assumed.
The important quantities of the $B^0_d$ system, such as $\sin 2 \beta$ and $\Delta m_d$, 
are determined under this assumption. However, the meson system interacts with its
environment. This interaction can lead to decoherence in the  mesons
even before they decay. In our formalism this decoherence is modelled by a single 
parameter $\lambda$. It is desirable to re-examine the procedures of determination of 
$\sin 2 \beta$ and $\Delta m_d$ in meson systems with decoherence.
We find that the present values of these two quantities are modulated by $\lambda$. 
Re-analysis of $B^0_d$ data from B-factories and LHCb can lead to a clean determination 
of $\lambda$, $\sin 2 \beta$ and $\Delta m_d$.
\end{abstract}

\maketitle 

\section{Introduction}
In neutral meson systems, quantum coherence plays a crucial role in the 
determination of many observables. However, any real system interacts with 
its environment and this interaction can lead to a loss of quantum coherence. 
The environmental effects may arise at a fundamental level, such as the 
fluctuations in a quantum gravity space-time background \cite{Hawking:1982dj,Ellis:1983jz}. 
They may also arise due to the detector environment itself.  Irrespective of 
the origin of the environment, its effect on the neutral meson systems can 
be taken into account by using the ideas of open quantum systems \cite{weiss, bp, bg03}. 
This formalism enables the inclusion of effects such as decoherence and dissipation in a systematic manner 
\cite{Banerjee:2014vga}.  Such an inclusion is
in accordance with the general principle of fluctuation-
dissipation theorem which states that dissipation is balanced by fluctuations.

The time evolution of neutral mesons, which are coherently produced in meson 
factories, are used to measure a number of parameters of the standard model 
of particle physics and also to search for physics beyond the standard model. 
However, decoherence is an unavoidable phenomenon as any physical system 
is inherently open due to its inescapable interactions with a pervasive environment.
With the inclusion of the decoherence effects, the measured values of some of 
these parameters can get masked. 
As the source of decoherence in the case of mesons could be expected 
to be coming from a much finer scale, it may happen that the numerical value of some of the 
masked observables are not greatly affected. This should however be verified experimentally.

In this work, we study the effect of decoherence 
on the important observables in the $B^0_d$ meson system, such as the CP 
violating parameter $\sin 2 \beta$ and the $B^0_d-\bar{B^0_d}$ mixing parameter 
$\Delta m_d$. We show that these parameters are affected by decoherence. 
So far only one attempt has been made to determine decoherence in
$B_d$ meson system \cite{Bertlmann:2001iw}. The bounds on the decoherence
parameter were obtained from the data on $R_d$, the ratio of the total
same-sign to opposite sign dilepton rates in the decays of coherent
$B_d - \bar{B}_d$ coming from the $\Upsilon (4S)$ decays. The data on
$R_d$ has not been updated in the last two decades \cite{R-chi}, whereas the
B-factories have provided direct and precise information on the
$B_d - \bar{B}_d$ mixing parameters.
In this work, we also suggest a number of methods which will enable clean determination 
of the decoherence parameter along with the other observables quite easily at the LHCb or B-factories.
We also attempt determination of the decoherence parameter and $\Delta m_d$ using 
Belle data on the time dependent flavor asymmetry of semi-leptonic $B^0_d$ decays as 
given in Ref.~\cite{Go:2007ww}.

The evolution of the $B_d^0$ system is built up from first principles. 
The effect of the environment forces the evolution 
to be a semi-group rather than a unitary one \cite{lindblad,sudarshan,Banerjee:2014vga}. 
We use the density matrix formalism to represent the time evolution of the $B^0_d$ system.
This ensures the complete positivity of the state of the system and hence its physical validity.
In this formalism, the decoherence is modelled by a single parameter $\lambda$.
By construction, the density matrices are trace preserving.

The work presented here, we hope, would lead to the inclusion of the 
effects of decoherence in the analysis of data from the $B^0_d$ systems.
It may be worthwhile to reanalyze the data from the B factories and LHCb 
to verify if a signature of decoherence is already inherent in it.  Thus a detailed study of $B_d^0$ observables 
can lead to tests of physics at scales much higher than those typical of flavour physics.

We first study the parameter $\sin 2 \beta$, whose measurement is the first signal for CP violation 
outside the neutral kaon system. The precision measurement of its value is the corner stone in
establishing the CKM mechanism for CP violation. With the inclusion of the decoherence effects,
it turns out that the experimentally measured CP asymmetry depends both on the decoherence parameter $\lambda$
and the angle $\beta$ of the unitarity triangle. Next we study $\Delta m_d$, which denotes the mixing
in the $B^0_d$ system and is an important input in extracting $\sin 2 \beta$ from the measured
time dependent CP asymmetry. We find that $\Delta m_d$ is also affected by 
the decoherence effects. Finally, we suggest a method of analysis by which the three quantities,
(a) $\lambda$, (b) $\Delta m_d$ and (c) $\sin 2 \beta$ can all be measured.

\section{Determination of $\sin 2 \beta$}
In the following, we develop the formalism which is applicable to $B^0_d$ as well 
as $B^0_s$ mesons. We are interested in the decays of $B^0$ and $\bar{B}^0$
mesons as well as $B^0 \leftrightarrow \bar{B}^0$ oscillations. To describe the 
time evolution of all these transitions, we need a basis of three states:
$\ket{B^0}$, $\ket{\bar{B}^0}$ and $\ket{0}$, where $\ket{0}$ 
reprents a state with no $B$ meson and is required for describing the decays.
In this basis, we can define $\rho_{B^0(\bar{B}^0)}(0)$, the initial 
density matrix for the state which starts out as $B^0(\bar{B}^0)$.
The time evolution of these matrices is governed by the Kraus operators
$K_i (t)$ as $\rho(t) = \sum_i K_i (t) \rho(0) K_i^\dagger (t)$ \cite{alle}.  
The Kraus operators are constructed taking into account the decoherence 
in the system which occurs due to the evolution under the influence of 
the environment \cite{sudar,alok}. The time dependent density matrices are
\bea
\frac{\rho_{B^0}(t)}{\frac{1}{2} e^{-  \Gamma t}}&=& 
\left(\begin{array}{ccc} 
 a_{ch} + e^{-  \lambda t} a_c &  - a_{sh} -i  e^{-  \lambda t} a_s & 0  \\ 
- a_{sh} + i e^{-  \lambda t} a_s & a_{ch} - e^{-  \lambda t} a_{c}  & 0 \\ 
0 &0   & 2 (e^{  \Gamma t}- a_{ch})
\end{array}\right),
\nonumber\\
\frac{\rho_{\bar{B^0}}(t)}{\frac{1}{2} e^{-  \Gamma t}}&=& 
\left(\begin{array}{ccc} 
 a_{ch} - e^{-  \lambda t} a_c &  - a_{sh} +i  e^{-  \lambda t} a_s & 0  \\ 
- a_{sh} - i e^{- \lambda t} a_s & a_{ch} + e^{-  \lambda t} a_{c}  & 0 \\ 
0 &0   & 2 (e^{ \Gamma t}- a_{ch})
\end{array}\right),
\label{dm-bbbar}
\eea
for $B^0$ and $\bar{B^0}$, respectively. In the above equation,  
$a_{ch}=\cosh \left(\frac{ \Delta \Gamma\, t}{2}\right)$, 
$a_{sh}=\sinh \left(\frac{ \Delta \Gamma\, t}{2}\right)$,
$a_c=\cos\left(\Delta m \,t\right)$,  $a_s=\sin\left(\Delta m\, t\right)$, 
$\Gamma=(\Gamma_L + \Gamma_H)/2$, $\Delta \Gamma = \Gamma_L -\Gamma_H$,  
where $\Gamma_L$ and $\Gamma_H$ are the respective decay widths of the 
decay eigenstates $B^0_L$ and $B^0_H$. Also $\lambda$ is the decoherence parameter, 
due to the interaction between one-particle system and 
its environment. As our main motivation is to bring out the fact that fundamental parameters
of $B-\bar{B}$ mixing and B sector CP violation are affected by
decoherence, here, and from here on, we will neglect the small mixing induced CP violation
to keep our formulae simple.
 
We define the decay amplitudes $A_f \equiv A(B^0 \to f)$ and 
$\bar{A}_f \equiv A(\bar{B}^0 \to f)$. The hermitian operator describing
the decays of the $B^0$ and $\bar{B}^0$ mesons into $f$ is 
\be 
\mathcal{O}_f = \left(\begin{array}{ccc} 
|A_f|^2 & {A_f}^* \bar{A}_f  & 0  \\ 
  A_f {\bar{A}_f}^* &  |\bar{A}_f|^2& 0 \\ 
0 &0   & 0
\end{array}\right).
\label{of}
\ee
The probability, $P_f(B^0/\bar{B}^0;t)$, of an initial $B^0/\bar{B}^0$ 
decaying into the state $f$ at time $t$ is given by ${\rm Tr}\left[{\cal O}_f 
\,\rho_{B^0(\bar{B}^0)(t)}\right]$.

Let us now consider $B^0_d \to J/\psi K_S$ decay.  One can  define a CP violating observable
\be 
{\mathcal{A}}_{J/\psi K_S} (t) = \frac{P_{J/\psi K_S}(\bar{B^0_d };t)- P_{J/\psi K_S}(B^0_d ;t)}
{P_{J/\psi K_S}(\bar{B^0_d };t)+ P_{J/\psi K_S}(B^0_d ;t)}\,.
\ee
Calculating the probabilities using Eqs.~(\ref{dm-bbbar}) and (\ref{of}) we get
\be 
\frac{{\mathcal{A}}_{J/\psi K_S} (t)}{e^{-  \lambda t}}= 
\frac{\left(|\lambda_f|^2-1\right) \cos\left(\Delta m_d t\right)+ 2 {\rm Im}(\lambda_f) \sin\left(\Delta m_d t\right)}
{\left(1+|\lambda_f|^2\right) \cosh \left(\frac{ \Delta \Gamma_d t}{2}\right)-2{\rm Re}(\lambda_f)
\sinh \left(\frac{ \Delta \Gamma_d t}{2}\right)},
\label{cpasy}
\ee
where
$\lambda_f= A(\bar{B^0_d} \to J/\psi K_S))/ A(B^0_d \to J/\psi K_S)$. 
Putting $\lambda=0$ in the above equation, we get the usual expression 
for CP asymmetry in the interference of mixing and decay. Thus the presence 
of decoherence modifies the expression for CP asymmetry  in the interference 
of mixing and decay.

In order to determine $\sin 2\beta$ from asymmetry defined in Eq.~\ref{cpasy},  it is usually assumed that, $\Delta \Gamma_d \approx 0$, $|\lambda_{f}|=1$, i.e., no direct CP asymmetry and ${\rm Im}(\lambda_f) \approx \sin 2\beta$. With these approximations,
the above expression simplifies to 
\be
{\mathcal{A}}_{J/\psi K_S} (t) = \sin 2\beta\, e^{- \lambda t } \sin\left(\Delta m_d \,t\right)\,.
\label{tdcpasym} 
\ee
Therefore we see that the coefficient of $\sin\left(\Delta m_d \, t\right)$ 
in the CP asymmetry is $\sin 2\beta\, e^{-  \lambda t}$ and not
$\sin 2\beta$! The measurement of $\sin 2 \beta$ is masked by the presence of
decoherence. Thus in order to have a clean determination of $\sin 2\beta$, 
an understanding of $\lambda$ is imperative.

Decoherence is expected to come from a scale much finer than that
of flavor physics and is likely to be small. Therefore, in the actual
comparison to the data, one should include all the known effects, which
are usually neglected in the extraction of $\sin 2 \beta$ and then do
a fit for clean determination of $\sin 2 \beta$ and  $\lambda$. The full fledged formula, of course, will
include the CP violation in mixing and decay width $\Delta \Gamma_d$.  Apart from these effects, one should also take into 
account the penguin contributions. The theoretical precision for the extraction of CP violating phase $\sin 2 \beta$ from the 
CP asymmetry of $B^0_d \to J/\psi K_S$ decay, defined in Eq.~\ref{cpasy}, is limited by  contributions from doubly 
Cabibbo-suppressed penguin topologies \cite{ Fleischer:1999zi,DeBruyn:2014oga}. This involves computation of
non-perturbative  hadronic parameters which, at present, cannot be achieved reliably using QCD. However, a way 
to control the penguin effects is offered by the $U$-spin symmetry of strong interactions 
which relates $B^0_s \to J/\psi K_S$ to $B^0_d \to J/\psi K_S$ \cite{Fleischer:1999nz}.
Ref.~\cite{DeBruyn:2014oga} discusses the constraining of the relevant penguin parameters by making use of this symmetry
as well as plausible assumptions for various modes of similar decay dynamics.

\section{Determination of $\Delta m_d$}
 It is obvious that in order to determine $\sin 2\beta$, we need to know 
$\Delta m_d$ and $\lambda$. If $\Delta m_d$ is measured using observables 
which are independent of $\lambda$, then we only  need to determine $\lambda$ 
for the clean extraction of $\sin 2 \beta$. If the determination of $\Delta m_d$ 
is also masked by the presence of decoherence then we need to have a clean 
determination of $\Delta m_d$. 

The present world average of $\Delta m_d$ quoted 
in PDG is $(0.510 \pm 0.003)\,\rm ps^{-1}$ \cite{Agashe:2014kda} which is an average of measurements of 
$\Delta m_d$ from  OPAL \cite{opal}, ALEPH \cite{aleph}, DELPHI \cite{delphi}, 
L3 \cite{l3}, CDF \cite{cdf},  BaBar \cite{babar}, Belle \cite{belle}, D0 
\cite{do} and LHCb \cite{lhcb} experiments. There are several ways in which 
$\Delta m_d$ can be determined experimentally. LHCb, CDF and D0 experiments 
determine $\Delta m_d$ by measuring rates that a state that is pure $B^0_d$ at 
time $t=0$, decays as either as $B^0_d$ or $\bar{B^0_d}$ as function of proper 
decay time. In the presence of decoherence, the survival (oscillation) probability 
of initial $B^0_d$ meson to decay as $B^0_d (\bar{B}_d^0)$ at a proper decay time $t$ is given by
\be 
P_{\pm} (t,\lambda) = \frac{e^{-\Gamma t}}{2}\left[\cosh(\Delta \Gamma_d t/2) \pm e^{-\lambda t} \cos (\Delta m_d t)\right]\,.
\label{spm}
\ee
The positive sign applies when the  $B^0_d$ meson decays with the same flavor as its production and 
the negative sign when the particle decays with opposite flavor to its production.
We see that the survival (oscillation) probability of $B^0_d$ is $\lambda$ dependent! 
The time dependent mixing asymmetry, used to determine $\Delta m_d$, is then given by
\be 
A_{\rm mix} (t,\lambda)=\frac{P_{+} (t,\lambda)-P_{-} (t,\lambda)}{P_{+} (t,\lambda)+P_{-} (t,\lambda)} 
=  e^{-\lambda t} \frac{\cos (\Delta m_d t)}{\cosh (\Delta \Gamma_d t/2)}\,. 
\label{amix}
\ee
Thus we see that the in the limit of neglecting $\Delta \Gamma_d$, the otherwise pure cosine dependence of mixing asymmetry is modulated by $e^{-\lambda t}$. Belle and BaBar experiments determine $\Delta m_d$ by measuring time 
dependent probability $P_+(t)$ of observing unoscillated $B^0_d \bar{B^0_d}$ events and 
$P_-(t)$ of observing oscillated $B^0_d B^0_d$/$\bar{B^0_d} \bar{B^0_d}$ events for two 
neutral $B_d$ mesons produced in an entangled state in the decay of the $\Upsilon(4S)$ 
resonance. The expressions for $P_\pm (t)$, in the presence of decoherence, are the same 
as those given in Eq.~(\ref{spm}), except that the proper time $t$ is replaced by the 
proper decay-time difference $\Delta t$ between the decays of the two neutral $B_d$ mesons. 
Therefore, we see that the determination of $\Delta m_d$ at LHCb, CDF, D0, Belle and BaBar 
experiments is masked by the presence of $\lambda$. The true value of $\Delta m_d$, along with $\Delta \Gamma_d$, 
can be determined by a three parameter ($\Delta m_d,\, \Delta \Gamma_d,\,\lambda$) fit to the time 
dependent mixing asymmetry $A_{\rm mix} (t,\lambda)$ defined in Eq.~(\ref{amix}). 
This in turn will enable a determination of true value of $\sin 2 \beta$ using Eq.~(\ref{cpasy}).

Determination of $\Delta m_d$ in the LEP experiments is mainly based on time independent measurements, i.e., 
from the ratio of the total same-sign to opposite-sign semileptonic rates ($R_d$) or the total $B^0_d-\bar{B}^0_d$ mixing 
probability ($\chi_d$).  We shall now see  that these observables are also $\lambda$ dependent. Therefore all the 
methods used to determine $\Delta m_d$ depend upon $\lambda$.

\section{Correlated $B^0_d$ meson semi-leptonic decays}
The entangled $B^0_d-\bar{B^0_d}$ mesons, produced in the decay of the $\Upsilon(4S)$ 
resonance, can both decay semi-leptonically. The effects of decoherence
on the resulting dilepton signal was studied in \cite{Bertlmann:2001iw}. 
Here we calculate these effects using the formalism described in the previous section. 
The entangled $B^0_d-\bar{B}_d^0$ state can be written as
\be
\ket{\psi(0)} =\frac{1}{\sqrt{2}} \left(\ket{B_d \bar{B^0_d}}-\ket{\bar{B^0_d} B_d}\right)\,.
\ee
The time evolution of the above state is described by the following density matrix 
\cite{Huet:1994kr,Benatti:1999cq,Benatti:2001tv}:
\bea
\rho(t_1,t_2)&=&\frac{1}{2} \Big(\rho_1(t_1)\otimes \rho_2(t_2) + \rho_2(t_1)\otimes \rho_1(t_2)\nonumber \\
&&- \rho_3(t_1)\otimes \rho_4(t_2)-\rho_4(t_1)\otimes \rho_3(t_2)\Big)\,,
\eea
where $\rho_1(t)=\rho_{B^0}(t)$, $\rho_2(t)=\rho_{\bar{B^0}}(t)$ which are given in Eq.~(\ref{dm-bbbar}),
while $\rho_{3/4}(t) = \sum_i K_i \rho_{3/4}(0) K_i^{\dagger}$, where $\rho_{3/4}(0) = \ket{B^0(\bar{B^0})}
\bra{\bar{B^0} (B^0)}$ and are given by
\bea
\frac{\rho_3(t)}{\frac{1}{2} e^{-  \Gamma t}}&=& 
\left(\begin{array}{ccc} 
 -a_{sh} -i e^{-  \lambda t} a_s &  a_{ch} + e^{-  \lambda t} a_c & 0  \\ 
a_{ch} - e^{-  \lambda t} a_c & -a_{sh} +i e^{-  \lambda t} a_s & 0 \\ 
0 &0   & 2 a_{sh}
\end{array}\right),
\nonumber\\
\frac{\rho_4(t)}{\frac{1}{2} e^{-  \Gamma t}}&=&
\left(\begin{array}{ccc} 
-a_{sh} +i e^{-  \lambda t} a_s &  a_{ch} - e^{-  \lambda t} a_c & 0  \\ 
a_{ch} + e^{-  \lambda t} a_c & -a_{sh} -i e^{-  \lambda t} a_s   & 0 \\ 
0 &0   & 2 a_{sh}
\end{array}\right).
\eea
Here the parameters are as in Eq.~(\ref{dm-bbbar}). The double decay rate, $G(f,t_1;g,t_2)$, 
that the left-moving meson decays at proper time $t_1$ into a final state $f$,
while the right-moving meson decays at proper time $t_2$ into the final state $g$,  
is then given by ${\rm Tr}\,[ (\mathcal{O}_f \otimes \mathcal{O}_g ) \, \rho(t_1,t_2)]$. 
From this a very useful quantity called the single time distribution, $\Gamma(f,g;t)$, can be defined as 
 $\Gamma(f,g;t)=\int_0^{\infty} d\tau\, G(f,\tau+t;g,\tau)$, where $t=t_1-t_2$ is taken to be positive.

We now  consider the decays of $B^0_d$ mesons into semileptonic states $h\, l\, \nu$, where $h$ stands for any allowed 
charged hadronic state. Under the assumption of CPT conservation and no violation of
$\Delta B = \Delta Q$ rule, the amplitudes for $B^0_d/\bar{B^0_d}$ into $h^- l^+ \nu$ can be written as 
\be 
A \left(B^0_d \to h^- l^+ \nu \right)=M_h\,, \qquad A \left(\bar{B^0_d} \to h^- l^+ \nu \right)=0\,,
\label{ahm}
\ee
whereas the amplitudes for $B^0_d/\bar{B^0_d}$ into $h^+ l^- \bar{\nu}$ are 
\be 
A\left(B^0_d \to h^+ l^- \bar{\nu} \right)=0\,, \qquad A\left(\bar{B^0_d} \to h^+ l^- \bar{\nu} \right)=M^*_h\,.
\label{ahp}
\ee
There are two important observables which can be affected by interaction with the environment.
 One is the ratio of the total same-sign to opposite-sign semileptonic rates
 \be
 R_d = \frac{\Gamma(h^+,h^+)+\Gamma(h^-,h^-)}{\Gamma(h^+,h^-)+\Gamma(h^-,h^+)}\,,
 \ee
 and the other is the total $B^0_d-\bar{B^0_d}$ mixing probability
 \be
 \chi_d = \frac{\Gamma(h^+,h^+)+\Gamma(h^-,h^-)}{\Gamma(h^+,h^+)+\Gamma(h^-,h^-)+\Gamma(h^+,h^-)+\Gamma(h^-,h^+)}\,.
 \ee
Time independent probabilities, $\Gamma(f,g)$, can be obtained by integrating 
the distribution $\Gamma(f,g;t)$ over time.

 The expressions for $R_d$ and $\chi_d$ are obtained to be 
 \bea
 R_d &=& \frac{1-(1-y^2)\left( (1+\lambda')^2 +x^2\right)^{-1} } 
 {1+(1-y^2)\left( (1+\lambda')^2 +x^2\right)^{-1} }\,,
 \label{Rd}\\
 \chi_d &=& \frac{1}{2} \left[1-(1-y^2)\left( (1+\lambda')^2 +x^2\right)^{-1} \right]\,,
 \label{chid}
 \eea
where we $x=\Delta m/\Gamma$, $y=\Delta \Gamma/2\Gamma$ and $\lambda'=\lambda/\Gamma$. 
 We see that  $R_d$ and $\chi_d$ are both functions of $(1-y^2)$ and $(1+\lambda')^2$.
It is interesting to note that in the limit of small $\lambda'$ and $y$, these combinations have a linear term in $\lambda'$ but only a quadratic term in $y$.
Thus we see that along with $\Delta m_d$ and $\Delta \Gamma_d$, 
these observables also depend upon the decoherence parameter $\lambda$.
 
For the observable $R_d$, the last experimental update was given about two decades ago \cite{R-chi}. 
This value was used in ref. \cite{Bertlmann:2001iw} to estimate the value of $\lambda$ to 
be $(-0.072\pm 0.118)$ $\rm ps^{-1}$.  It is 
important to reanalyze the BaBar and Belle data on the time dependent mixing asymmetry in terms of 
the three parameters $(\lambda, \Delta m_d, \Delta \Gamma_d)$ using the expression given in Eq.~(\ref{amix}). 
One should also obtain the  value of $\chi_d$ from CDF, DO and LHCb. Then the expression in 
Eq.~(\ref{chid}) can be verified using the values obtained from
the fit to the time dependent mixing asymmetry. This will provide an additional 
consistency check on assumptions made regarding decoherence. Finally, the 
values of $\lambda$, $\Delta m_d$ and $\Delta \Gamma_d$ from the $A_{\rm mix}(t, \lambda)$
fit can be used in Eq.~(\ref{cpasy}) to obtain a clean measurement of
$\sin 2 \beta$.

 The present analysis can easily be extended to the $B^0_s$ system as well. 
The expression for the time dependent 
CP asymmetry in the mode $B^0_s \to J/\psi \phi$ will be a function of four parameters: $\lambda$, $\sin 2 \beta_s$, 
$\Delta m_s$ and $\Delta \Gamma_s$. The time dependent 
mixing asymmetry defined in Eq.~(\ref{amix}) will determine $\lambda$, $\Delta m_s$ and $\Delta 
\Gamma_s$. These two time-dependent asymmetries should be re-analysed using 
a four parameter fit for a clean  determination of $\sin 2 \beta_s$, $\Delta m_s$, $\Delta \Gamma_s$ and $\lambda$.
Also, like $\sin 2 \beta_d$, the extraction of $\sin 2 \beta_s$ from time dependent 
CP asymmetry in the mode $B^0_s \to J/\psi \phi$ is restricted due to penguin pollution. 
In this case, the analysis of CP violation is more involved in comparison to $B^0_d \to J/\psi K_S$. 
This is due to the fact that the final state involves two vector mesons. The admixture of different CP
eigenstates can be disentangled through a time-dependent angular analysis of the decay products of the vector mesons \cite{Dighe:1995pd,Dighe:1998vk}. 
The penguin contribution to $B^0_s \to J/\psi \phi$ can be estimated using decays $B^0_d \to J/\psi \rho$ and $B^0_s \to J/\psi \bar{K^*}$ 
\cite{Fleischer:1999zi,Faller:2008gt}.

\section{Estimation of $\lambda$: An Example}

Here we make an attempt of a clean determination of $\lambda$, $\Delta m_d$ and $\Delta \Gamma_d$ 
using the experimental data of the time dependent flavor asymmetry of semi-leptonic $B^0_d$ decays as 
given in Ref.~\cite{Go:2007ww}. We perform a $\chi^2$ fit to $A_{\rm mix} (\Delta t,\lambda)$, 
using the efficiency corrected distributions given in Table I of Ref.~\cite{Go:2007ww}. 
First, the fit is done by assuming no decoherence, i.e., $\lambda=0$. In this case, we find $\Delta m_d=(0.489\pm 0.010)$ $\rm ps^{-1}$ 
and $\Delta \Gamma_d = (0.087\pm 0.054)$ $\rm ps^{-1}$  with $\chi^2/d.o.f=8.42/9$. We then redo the fit including decoherence. 
This gives $\lambda=(-0.012\pm 0.019)$ $\rm ps^{-1}$  along with $\Delta m_d=(0.490\pm 0.010)$ $\rm ps^{-1}$ 
and $\Delta \Gamma_d = (0.144\pm0.088)$ $\rm ps^{-1}$  with $\chi^2/d.o.f=8.02/8$. 
Thus we see that the decoherence parameter $\lambda$ is very loosely bounded. The upper limit on $\lambda$ is 
$0.03$ $\rm ps^{-1}$ at 95\% C.L.  We also find in this example that $\Delta m_d$ is numerically unaffected where 
as $\Delta \Gamma_d$ can be affected by inclusion of decoherence.
Given the wealth of data coming from LHCb and expected from the KEK Super B factory, a clear picture is expected to emerge.

\section{Conclusions} 
In this work, we have studied the effect of decoherence on two important 
observables $\sin 2 \beta$ and $\Delta m_d$ in a neutral meson 
system. We find that the asymmetries which determine these quantities are 
also functions of the decoherence parameter $\lambda$. Hence 
it is imperative to measure $\lambda$ for a clean determination of these quantities. We suggest 
a re-analysis of the data on the above asymmetries for an accurate measurement of all the three
quantities $\lambda$, $\sin 2 \beta$ and $\Delta m_d$.
The present analysis can easily be extended to the $B^0_s$ system as well.

{\em Acknowledgments.\textemdash} We thank Kajari Mazumdar for helpful 
discussions on several parts of this analysis. 
We thank David Hitlin for suggesting an example for determination of $\lambda$.
We also thank Joaquim Matias, David London and B. Ananthanarayan for useful 
comments. The work of AKA and SB is supported by CSIR, Government of India, grant no: 03(1255)/12/EMR-II.



\begin{thebibliography}{10}

 \bibitem{Hawking:1982dj}
  S.~W.~Hawking,
  Commun.\ Math.\ Phys.\  {\bf 87} (1982) 395.
  
  \bibitem{Ellis:1983jz} 
  J.~R.~Ellis, J.~S.~Hagelin, D.~V.~Nanopoulos and M.~Srednicki,
  Nucl.\ Phys.\ B {\bf 241}, 381 (1984).
 
 \bibitem{weiss} U. Weiss, \textit{Quantum Dissipative Systems, Third
 Edition} (World Scientific 2008). 

 \bibitem{bp} H.-P. Breuer and F. Petruccione, \textit{The Theory
 of Open Quantum Systems} (Oxford University Press 2002).

 \bibitem{bg03}  S. Banerjee  and R. Ghosh, Phys. \ Rev.\  E {\bf 67}, 056120 (2003).

 \bibitem{Banerjee:2014vga} 
  S.~Banerjee, A.~K.~Alok and R.~MacKenzie,
  arXiv:1409.1034 [hep-ph].
  
 
 \bibitem{Bertlmann:2001iw} 
  R.~A.~Bertlmann and W.~Grimus,
  Phys.\ Rev.\ D {\bf 64}, 056004 (2001)
  [hep-ph/0101160].
  
  \bibitem{R-chi} 
  H.~Albrecht {\it et al.}  [ARGUS Collaboration],
  Phys.\ Lett.\ B {\bf 324}, 249 (1994);
   J.~E.~Bartelt {\it et al.}  [CLEO Collaboration],
  Phys.\ Rev.\ Lett.\  {\bf 71}, 1680 (1993).
  
  \bibitem{Go:2007ww} 
  A.~Go {\it et al.}  [Belle Collaboration],
  Phys.\ Rev.\ Lett.\  {\bf 99}, 131802 (2007)
  [quant-ph/0702267 [QUANT-PH]].

 \bibitem{lindblad}
 G. Lindblad,  Commun. Math. Phys. {\bf 48}, 119 (1976).

 \bibitem{sudarshan}
 V. Gorini, A. Kossakowski and E. C. G. Sudarshan, J. Math. Phys. {\bf 17}, 821 (1976).

 \bibitem{alle} R. Alicki and K. Lendi, \textit{Quantum Dynamical Semigroups and Applications} (Lect. Notes
 Phys. 717 (Springer, Berlin Heidelberg 2007)).
 
  \bibitem{sudar} E. C. G. Sudarshan, P. M. Mathews and J. Rau, Phys. Rev. {\bf 121}, 920 (1961);
 K. Kraus, \textit{States, Effects and Operations: Fundamental Notions of Quantum Theory} (Springer Verlag 1983).

 \bibitem{alok} A. K. Alok {\em et al.}, Work in progress.


  \bibitem{Fleischer:1999zi} 
  R.~Fleischer,
  Phys.\ Rev.\ D {\bf 60}, 073008 (1999)
  [hep-ph/9903540].
  
  \bibitem{DeBruyn:2014oga} 
  K.~De Bruyn and R.~Fleischer,
  JHEP {\bf 1503}, 145 (2015)
  [arXiv:1412.6834 [hep-ph]].
  
  \bibitem{Fleischer:1999nz} 
  R.~Fleischer,
  Eur.\ Phys.\ J.\ C {\bf 10}, 299 (1999)
  [hep-ph/9903455].
  
  
  
\bibitem{Agashe:2014kda} 
  K.~A.~Olive {\it et al.}  [Particle Data Group Collaboration],
  Chin.\ Phys.\ C {\bf 38}, 090001 (2014).
  
  \bibitem{opal}
  G.~Abbiendi {\it et al.}  [OPAL Collaboration],
  Phys.\ Lett.\ B {\bf 493}, 266 (2000)
  [hep-ex/0010013].
  
  \bibitem{aleph}
    D.~Buskulic {\it et al.}  [ALEPH Collaboration],
  Z.\ Phys.\ C {\bf 75}, 397 (1997).
  
  \bibitem{delphi}
    J.~Abdallah {\it et al.}  [DELPHI Collaboration],
  Eur.\ Phys.\ J.\ C {\bf 28}, 155 (2003)
  [hep-ex/0303032].
  
  \bibitem{l3}
    M.~Acciarri {\it et al.}  [L3 Collaboration],
  Eur.\ Phys.\ J.\ C {\bf 5}, 195 (1998).
  
  \bibitem{cdf}
    F.~Abe {\it et al.}  [CDF Collaboration],
  Phys.\ Rev.\ D {\bf 60}, 072003 (1999)
  [hep-ex/9903011].
  
  \bibitem{babar}
    B.~Aubert {\it et al.}  [BaBar Collaboration],
  Phys.\ Rev.\ D {\bf 73}, 012004 (2006)
  [hep-ex/0507054].
  
  \bibitem{belle}
    K.~Abe {\it et al.}  [BELLE Collaboration],
  Phys.\ Rev.\ D {\bf 71}, 072003 (2005)
  [Erratum-ibid.\ D {\bf 71}, 079903 (2005)]
  [hep-ex/0408111].

  \bibitem{do}
    V.~M.~Abazov {\it et al.}  [D0 Collaboration],
  Phys.\ Rev.\ D {\bf 74}, 112002 (2006)
  [hep-ex/0609034].
  
  \bibitem{lhcb}
    R.~Aaij {\it et al.}  [LHCb Collaboration],
  Phys.\ Lett.\ B {\bf 709}, 177 (2012)
  [arXiv:1112.4311 [hep-ex]]; 
  R.~Aaij {\it et al.}  [LHCb Collaboration],
  Eur.\ Phys.\ J.\ C {\bf 73}, no. 12, 2655 (2013)
  [arXiv:1308.1302 [hep-ex]];
    R.~Aaij {\it et al.}  [LHCb Collaboration],
  Phys.\ Lett.\ B {\bf 719}, 318 (2013)
  [arXiv:1210.6750 [hep-ex]].
  
    \bibitem{Huet:1994kr} 
  P.~Huet and M.~E.~Peskin,
  Nucl.\ Phys.\ B {\bf 434}, 3 (1995)
  [hep-ph/9403257].
  
  \bibitem{Benatti:1999cq} 
  F.~Benatti and R.~Floreanini,
  Phys.\ Lett.\ B {\bf 465}, 260 (1999)
  [hep-ph/9909361].
  
  \bibitem{Benatti:2001tv} 
  F.~Benatti, R.~Floreanini and R.~Romano,
  Nucl.\ Phys.\ B {\bf 602}, 541 (2001)
  [hep-ph/0103239].
  
  \bibitem{Dighe:1995pd} 
  A.~S.~Dighe, I.~Dunietz, H.~J.~Lipkin and J.~L.~Rosner,
  Phys.\ Lett.\ B {\bf 369}, 144 (1996)
  [hep-ph/9511363].
  
  \bibitem{Dighe:1998vk} 
  A.~S.~Dighe, I.~Dunietz and R.~Fleischer,
  Eur.\ Phys.\ J.\ C {\bf 6}, 647 (1999)
  [hep-ph/9804253].
  
  \bibitem{Faller:2008gt} 
  S.~Faller, R.~Fleischer and T.~Mannel,
  Phys.\ Rev.\ D {\bf 79}, 014005 (2009)
  [arXiv:0810.4248 [hep-ph]].
  
  
\end{thebibliography}
\end{document}